\documentclass[prf,aps,preprint,nofootinbib]{revtex4-1}
\usepackage{graphicx}
\usepackage{amsfonts}
\usepackage{amssymb}
\usepackage{amsmath}
\usepackage{psfrag}
\usepackage{natbib}
\usepackage{mathtools}
\usepackage{subfigure,float}
\usepackage[colorlinks=true, linkcolor=blue, urlcolor=blue, citecolor=black]{hyperref}
\newcommand{\beqa}{\begin{eqnarray}}
\newcommand{\eeqa}{\end{eqnarray}}
\newcommand{\beq}{\begin{equation}}
\newcommand{\eeq}{\end{equation}}

\usepackage{tabularx}

\usepackage[usenames,dvipsnames]{color}
\usepackage[normalem]{ulem}

\definecolor{chcolor}{rgb}{0,0.4,0.6}

\begin{document}
\title{Soft wetting with (a)symmetric Shuttleworth effect}

\author{C. Henkel$^{1\star}$, M. H. Essink$^{2\star}$, Tuong Hoang$^2$, G. J. van Zwieten$^3$, E. H. van Brummelen$^4$, U. Thiele$^{1,5}$, J. H. Snoeijer$^2$ \\
}
\affiliation{$^1$ Institut f\"ur Theoretische Physik, Westf\"alische Wilhelms-Universit\"at M\"unster, Wilhelm-Klemm-Str.\ 9, 48149 M\"unster, Germany\\
$^2$Physics of Fluids Group, Faculty of Science and Technology, Mesa+ Institute, University of Twente, 7500 AE Enschede, The Netherlands \\
$^3$Evalf Computing, Burgwal 45, 2611 GG Delft, The Netherlands\\
$^4$Multiscale Engineering Fluid Dynamics Group, Department of Mechanical Engineering, Eindhoven University of Technology, P.O. Box 513, 5600 MB Eindhoven, The Netherlands\\
$^5$Center for Nonlinear Science (CeNoS), Westf{\"a}lische Wilhelms-Universit\"at M\"unster, Corrensstr.\ 2, 48149 M\"unster, Germany\\
$^\star$ These authors have equally contributed to this work.
}


\begin{abstract}
The wetting of soft polymer substrates brings in multiple complexities as compared to the wetting on rigid substrates. The contact angle of the liquid is no longer governed by Young's law, but is affected by the substrate's bulk and surface deformations. On top of that, elastic interfaces exhibit a surface energy that depends on how much they are stretched -- a feature known as the Shuttleworth effect (or as surface-elasticity). Here we present two models by which we explore the wetting of drops in the presence of a strong Shuttleworth effect. The first model is macroscopic in character and consistently accounts for large deformations via a neo-Hookean elasticity. The second model is based on a mesoscopic description of wetting, using a reduced description of the substrate's elasticity. While the second model is more empirical in terms of the elasticity, it enables a gradient dynamics formulation for soft wetting dynamics. We provide a detailed comparison between the equilibrium states predicted by the two models, from which we deduce robust features of soft wetting in the presence of a strong Shuttleworth effect. Specifically, we show that the (a)symmetry of the Shuttleworth effect between the ``dry" and ``wet" states governs  horizontal deformations in the substrate. Our results are discussed in the light of recent experiments on the wettability of stretched substrates.
\end{abstract}
\maketitle


\section{Introduction}

Drops on elastic substrates represent a paradigmatic example of ``soft wetting'', where capillarity-induced elastic substrate deformations dramatically affect the static and dynamic wetting behaviour of partially and completely wetting liquids \cite{AnSn2020arfm}.  Recent work has shown that substrates made from cross-linked polymer networks offer versatile routes to manipulate contact angles of droplets \cite{StDu2012sm,StDuPRL2013,LWBD2014jfm,BoSD2014sm,DervauxLimat2015}, as well as their spreading dynamics \cite{CaGS1996n,LoAL1996lb,KarpNcom15,GKAS2020sm}, directed motion \cite{SCPW2013potnaos,BBJG2018sm,ZDNL2018pnasusa} and condensation \cite{SokulerLangmuir2010}.  However, the full richness of these phenomena is only beginning to emerge and at present even a  quantitative understanding of the behaviour of a single drop of nonvolatile simple liquid is not yet complete.  Key challenges lie in the intricate effects of solid surface tension, and how it affects the force balance near the static three-phase contact line, while dynamics involves viscoelasticity of the substrate and elastocapillary interactions between droplets \cite{AnSn2020arfm}.

The capillarity of soft solids introduces a major complication as compared to liquid interfaces. Namely, in general, one expects the surface free energy to depend on the surface strain. This is known as surface elasticity or the Shuttleworth effect~\cite{Shut1950ppsa,MDSA2012prl,WeAS2013sm,LCWD2018l,AnSn2016el,SJHD2017arcmp}. Therefore, one needs to distinguish the (scalar) surface energy from the (tensorial) surface tension, neither of which can be treated as a universal material constant \cite{SJHD2017arcmp,AnSn2020arfm}. The influence of strain-dependent surface tension was recently explored experimentally by measuring contact angles on stretched substrates \cite{XJBS2017nc,STSR2018nc,XuSD2018sm}, but the results were contradictory. On the theoretical side, the Shuttleworth effect is only beginning to be explored for soft amorphous materials~\cite{SnRA2018prl,PAKZ2020prx}, but so far work was restricted to isolated contact lines.

\begin{figure}[H]
\centerline{\includegraphics[width=\textwidth]{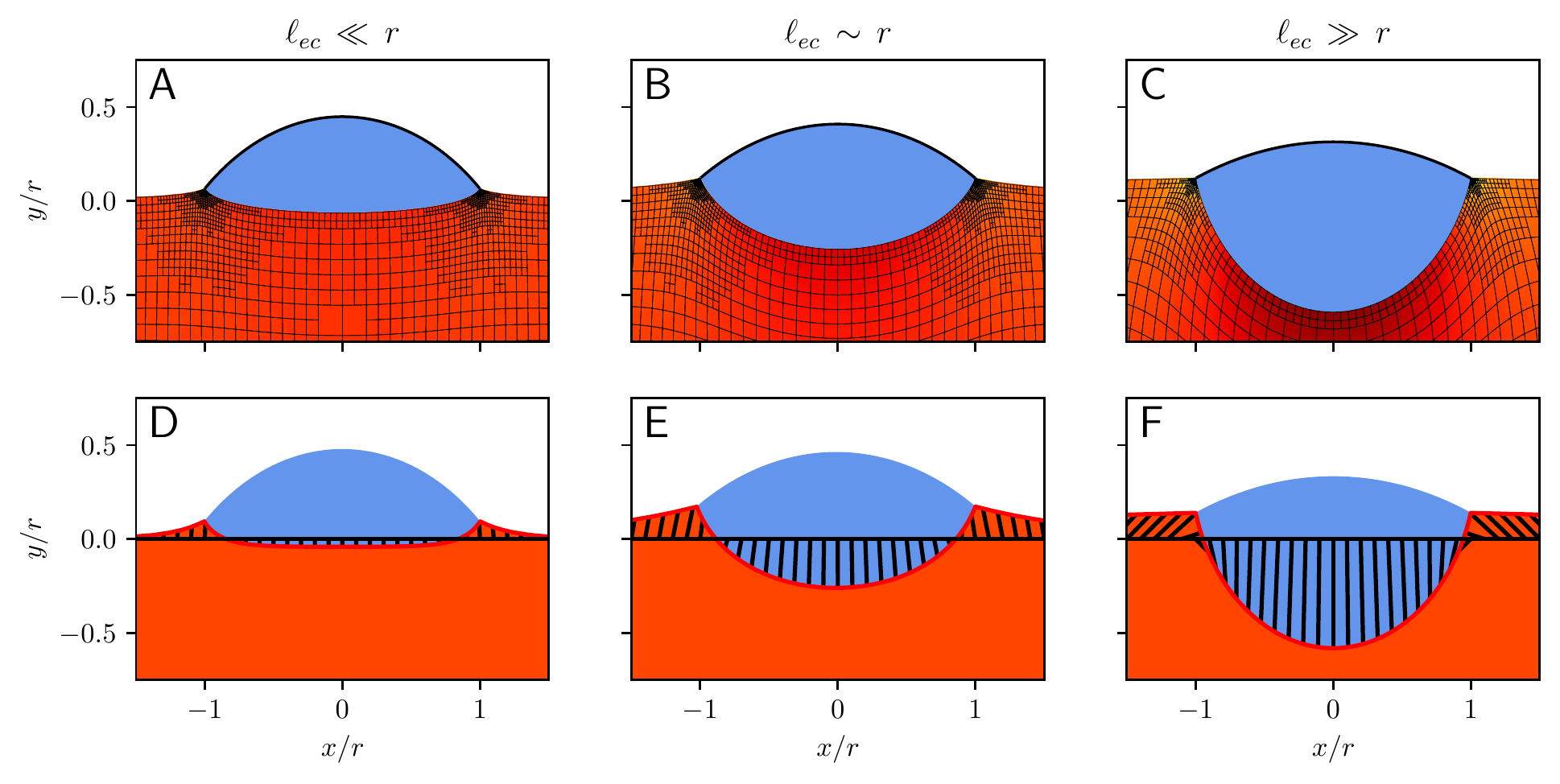}}
\caption{Drops on elastic substrates of decreasing stiffness, described using  two different modelling approaches: a macroscopic model based on a neo-Hookean bulk elasticity (top) and a mesoscopic gradient dynamics model using a Winkler foundation (bottom). In the top row (a-c) the substrate's elastic deformation is visible from the grids that in the reference state are straight horizontal/vertical. In the bottom row (d-f) the lines indicate the interface  displacement induced by the presence of the drop. Both models capture the transition from ``rigid" to ``liquid" wetting: the droplet sinks into the substrate and its liquid angle (with respect to the horizontal) decreases. This process is governed by the  elastocapillary length $\ell_{\rm ec}$ normalized by the drop radius $r$. Model parameters are Young's angle $\theta^0_Y=48.19^\circ$, Shuttleworth coefficients  $\gamma^1_{SV} = \gamma^1_{SL} = \gamma^0_{SL}$, and liquid contact angles (left) $\theta_L \approx 42^\circ$, (center) $\theta_L \approx 32^\circ$,  (right) $\theta_L \approx 22^\circ$.
}
\label{fig:sinking}
\end{figure}

Here we explore the static wetting of droplets on elastic substrates in the presence of the Shuttleworth effect (Fig.~\ref{fig:sinking}). We simultaneously present two modelling approaches, each of which has its own specific merits. On the one hand, we expand the macroscopic approach of Ref.~\cite{PAKZ2020prx}, which consistently accounts for large elastic deformation via a neo-Hookean elasticity in the presence of the Shuttleworth effect. The previous approach for single contact lines is now extended to droplets of finite volume (Fig.~\ref{fig:sinking}, top-row). 
On the other hand we extend the  mesoscopic thin-film model developed in Ref.~\cite{HeST2021sm}, where we now incorporate the Shuttleworth effect and allow for larger contact angles  (Fig.~\ref{fig:sinking}, bottom-row). The elasticity in this mesoscopic model is described using a reduced ``Winkler" foundation, which sacrifices some detail on the substrate's deformation but offers a great potential towards dynamical modelling of large ensembles of drops. We now show how the Shuttleworth effect can be introduced into the mesoscopic model, and offer a detailed comparison of the equilibrium states obtained with the macroscopic neo-Hookean model. This comparison includes the presentation of consistency conditions \cite{TSTJ2018l,HeST2021sm} that ensure the correct relation between macro- and mesoscale descriptions of wettability, in the presence of the Shuttleworth effect. As can already be inferred from Fig.~\ref{fig:sinking}, both models recover the ``sinking" of the drop into the substrate as the elastic modulus is decreased.

Our central finding is that the Shuttleworth effect has a major influence on the horizontal deformations of the substrate, while its effect on the normal displacements is relatively minor. Specifically, any asymmetry of the Shuttleworth effect between the ``dry" and ``wet" parts of the substrates induces large horizontal displacements. This is in line with previous predictions made in the rigid limit for very small deformations \cite{WeAS2013sm,AnSn2016el}, but now shown for arbitrary stiffness and for large deformations. In addition, we for the first time model the change of the liquid contact angle with stiffness in the presence of the Shuttleworth effect; again we find that Shuttleworth (a)symmetry is essential for the effective wettability.

The paper is organized as follows. In Sec.~\ref{sec:Seffect} we give a detailed description of the Shuttleworth effect. We develop both the Lagrangian formulation (common in solid mechanics) and the Eulerian formulation (common in fluid mechanics). Then, we develop the macroscopic and mesoscopic descriptions of wetting in Sec.~\ref{sec:macro}, where we address subtleties of contact lines in the presence of the Shuttleworth effect. Then, the two models are presented in Sec.~\ref{sec:twomodels} followed by the results in Sec.~\ref{sec:results}. The paper closes with a Discussion in Sec.~\ref{sec:discussion}, where we also sketch a perspective in terms of dynamics, showing how the mesoscopic model also allows the exploration of dynamical wetting in the presence of the Shuttleworth effect.

\section{The Shuttleworth effect:\\ Capillarity with a stretch-dependence} \label{sec:Seffect}

\subsection{Kinematics of surface stretch} 

\subsubsection{Lagrangian description}
\label{sec:lagrange}
\begin{figure}[H]
\centerline{\includegraphics[width=0.5\textwidth]{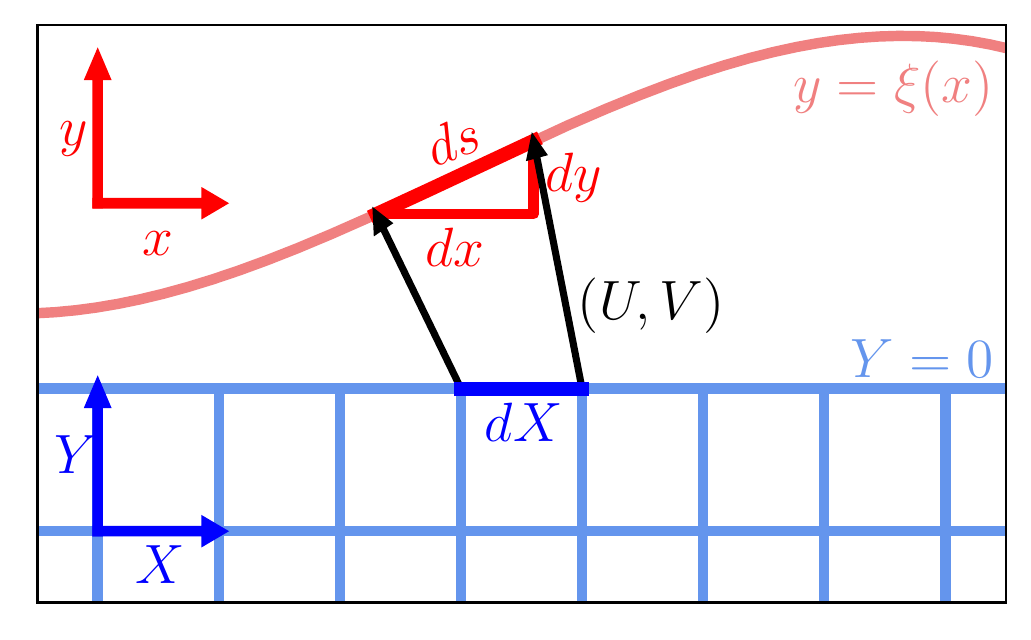}}
\caption{Substrate deformation defined by the mapping from material coordinates $\mathbf R = (X,Y)$ to current coordinates $\mathbf r = (x,y)$, which can be expressed via a displacement vector $(U,V)=\mathbf r - \mathbf R$. The free surface is defined as $Y=0$ and $y=\xi(x)$, respectively. The surface stretch $\lambda$ is defined as the ratio $ds/dX$.}
\label{fig:bla}
\end{figure}

Elastic deformations are described in terms of a mapping, where a point $\mathbf R$ in the reference configuration of the soft substrate (prior to deformation) is displaced to a point $\mathbf r$ in the current configuration (after deformation) \cite{MarsdenHughes1994,Holzapfel2000}. The mapping can be written as $\mathbf r = \chi(\mathbf R)$, where $\chi$ is called the deformation, which is assumed to be differentiable and invertible. As Fig.~\ref{fig:bla}, we focus on a substrate that is two-dimensional (assuming plain strain elasticity), so that its free surface is one-dimensional. This will facilitate a physical discussion in terms of scalar quantities,  avoiding the tensor algebra associated with two-dimensional manifolds. To be explicit, we employ Cartesian coordinates $\mathbf R=(X,Y)$ (also called ``material coordinates'') and $\mathbf r=(x,y)$ (also called ``current coordinates'' or ``spatial coordinates''), as indicated in Fig.~\ref{fig:bla}. The mapping can then be written as
\begin{eqnarray}
x &=& X + U(X,Y)\\
y &=& Y + V(X,Y).
\end{eqnarray}
where we introduced the horizontal and vertical displacements $U$ and $V$, respectively.

To facilitate the presentation, but without any essential restrictions, we now consider the free surface of the substrate in the reference configuration to be flat and to be located at $Y=0$. The relation of the lengths of a surface element in the reference configuration, $dX$, and the current deformed configuration, $ds$, then follows as 
\begin{equation}
ds^2 = dx^2 + dy^2 = \left[\left(\frac{\partial x}{\partial X} \right)^2 +  \left(\frac{\partial y}{\partial X} \right)^2\right] dX^2, \quad \mathrm{at} \quad Y=0.
\end{equation}
The ``surface stretch'' $\lambda$ is defined as the ratio of the surface measure in the deformed and undeformed configurations, i.e., 
\begin{equation}
\lambda^2 =  \left(\frac{\partial x}{\partial X} \right)^2 +  \left(\frac{\partial y}{\partial X} \right)^2  = 
 \left(1+ U' \right)^2 + V'^2 \quad \mathrm{at} \quad Y=0.
 \end{equation}
 This gives the ``Lagrangian definition'' of stretch, expressed in terms of functions that depend on the material coordinate $X$. 

\subsubsection{Eulerian description}
\label{sec:euler}
In fluid mechanics, capillarity is usually described using the shape of the interface, defined as $y=\xi(x)$ in Fig.~\ref{fig:bla}. Such a description is intrinsically ``Eulerian'' in nature, since it uses the current coordinate $x$ as a variable, and no allusion is made to any underlying material coordinate $X$. The length of a surface element is $ds=\sqrt{1+\xi'^2}\,dx$. However, in order to compute the surface stretch $\lambda$, we need to relate $ds$ to the original length $dX$ (see Fig.~\ref{fig:bla}).
This relation can be found by defining the \emph{inverse} mapping, $\mathbf R = \chi^{-1}(\mathbf r)$, or in Cartesian coordinates $X(x,y)$ and $Y(x,y)$. We remind the reader that the free surface is located at $Y=0$ (Lagrangian), or $y=\xi(x)$ (Eulerian). Evaluating the inverse mapping at the surface, we thus find 
 \begin{eqnarray}
X &=&  x - U\left(X(x,\xi(x)),0 \right) \equiv x - u(x)\label{eq:defu1}\\
Y &=&  0.\label{eq:defu2}
\end{eqnarray}
Here we introduced the horizontal displacement at the surface, $u(x)$, expressed as a function of the Eulerian coordinate $x$.

With these definitions in place, we can compute the original length of a surface element by taking the derivative of (\ref{eq:defu1}), giving $dX=(1-u')dx$. Combined with  $ds=\sqrt{1+\xi'^2}dx$ this gives the Eulerian definition of surface stretch:
\begin{equation}\label{eq:stretch}
\lambda =  \frac{ds}{dX}= \frac{\sqrt{1+\xi'^2}}{1-u'}.
\end{equation}
From this expression it is clear that one can change the material configuration of the substrate without changing its shape. Namely, even when the surface profile $\xi(x)$ is kept constant, one can vary the surface stretch upon changing $u'(x)$. 

\subsection{Surface energy, surface tension, surface-chemical potential}
We consider a soft solid with a free interface $\Omega_{SL}$ to a macroscopic liquid layer of thickness $h(x)$ that completely covers it and has itself a free surface $\Omega_{LV}$. The total capillary energy of the system reads 
\begin{equation}\label{eq:fcap}
\mathcal F_{\rm cap} = \int_{\Omega_{LV}} ds \, \gamma_{LV}  + \int_{\Omega_{SL}} ds \, \gamma_{SL}(\lambda) ,
\end{equation}
where $\gamma_{LV}$ and $\gamma_{SL}$ are the liquid-vapor and solid-liquid surface energy densities, respectively. Variation of the energy with respect to the substrate degrees of freedom gives rise to two distinct physical quantities: the surface tension $\Upsilon$ and the surface-chemical potential $\mu$~\cite{SnRA2018prl,AnSn2020arfm,PAKZ2020prx}. 
Here we show how these quantities emerge from the parameterization  based on $h(x),\xi(x)$, and $u(x)$, where $h(x)$ refers to the liquid-layer thickness. In terms of these functions, Eq.~(\ref{eq:fcap}) becomes
\begin{equation}\label{eq:fcapbis}
\mathcal F_{\rm cap}[h,\xi,u] = \int dx\,\left\{ m(h'+\xi')\,\gamma_{LV}  + m(\xi')\,\gamma_{SL}(\lambda) \right\}
\end{equation}
where we introduced metric factors $m(z)=\sqrt{1+z^2}$ for the two interfaces, facilitating a description of the problem on the $x$-domain. Note that $m'(z)=z/m(z)$.

The surface tension $\Upsilon$ and chemical potential $\mu$ indeed appear during the variations of
$\mathcal F_{\rm cap}$. We therefore present the functional derivatives,  keeping in mind that the final minimization scheme will include additional energies and Lagrange multipliers related to side conditions like fixed volume.
The functional derivative of (\ref{eq:fcapbis}) with respect to the liquid layer thickness, $h(x)$, gives
\begin{equation}\label{eq:delh}
\frac{\delta F_{\rm cap}}{\delta h} = - \gamma_{LV} \frac{\partial}{\partial x} \left(\frac{h'+\xi'}{m(h'+\xi')} \right).
\end{equation}
On the right hand side we can recognize the usual Laplace pressure; namely, working out the derivative with respect to $x$ gives the curvature of the liquid-vapor interface $(h''+\xi'')/m(h'+\xi')^3$. A similar result is obtained from the functional derivative of (\ref{eq:fcapbis}) with respect to the shape of the solid-liquid interface $\xi(x)$:
\begin{equation}\label{eq:delxi}
\frac{\delta F_{\rm cap}}{\delta \xi} = - \frac{\partial}{\partial x} \left( \Upsilon_{SL} \frac{\xi'}{m(\xi')} \right)
- \gamma_{LV} \frac{\partial}{\partial x} \left(\frac{h'+\xi'}{m(h'+\xi')} \right).
\end{equation}
An important difference with respect to the liquid-vapor interface is that this expression now involves the surface tension
\begin{equation}
\Upsilon_{SL} \equiv \gamma_{SL} + \lambda \frac{\partial \gamma_{SL}}{\partial \lambda} ,
\label{eq:surftens}
\end{equation}
which contains an extra term associated to the stretch-dependence, $\partial \gamma_{SL}/\partial \lambda$. This reflects the Shuttleworth effect and also is the reason why one needs to distinguish between surface energy $\gamma_{SL}$ and surface tension $\Upsilon_{SL}$. Another important feature is that $\Upsilon_{SL}$ is no longer constant and can not be pulled out of the $x$-derivative. The stretch-dependence of $\Upsilon_{SL}$ is similar to the dependency of surface tension on surfactant concentration for liquid surfaces covered by surfactant molecules \cite{ThAP2016prf,TSTJ2018l}. In consequence, in analogy to the solutal Marangoni effect \cite{NepomnyashchyVelardeColinet2002}, a gradient in local stretch $\lambda$ will give rise to a tangential Marangoni-like force \cite{PAKZ2020prx}.

We can change the ``material composition" of the substrate independently of the interface shape. This is achieved by varying the horizontal displacements $u(x)$ at constant $\xi(x)$. Taking the functional derivative with respect to $u(x)$, we obtain
\begin{equation}\label{eq:delu}
\frac{\delta F_{\rm cap}}{\delta u} = - \frac{\partial \mu_{SL}}{\partial x}
\end{equation}
where we define a surface-chemical potential
\begin{equation}\label{eq:mu}
\mu_{SL} \equiv \lambda^2  \frac{\partial \gamma_{SL}}{\partial \lambda}.
\end{equation}
related to the conservation of the material points at the surface of the elastic substrate. The surface-chemical potential $\mu_{SL}$ governs the composition of material points along the substrate's interface.\footnote{%
This can be directly seen when taking into account that the surface stretch $\lambda$ is inverse to the density $\rho_\mathrm{s}$ of the material points at the surface of the elastic layer. Namely, $\lambda=\rho_0/\rho_\mathrm{s}$ where $\rho_0$ is the constant reference surface density of the undeformed layer. Expressed in $\rho_\mathrm{s}$ we have $\mu_{SL}=-\rho_0\,\frac{\partial \gamma_{SL}}{\partial \rho_\mathrm{s}}$ and $\Upsilon_{SL} = \gamma_{SL} - \rho_\mathrm{s} \frac{\partial \gamma_{SL}}{\partial \rho_\mathrm{s}}$ instead of \eqref{eq:mu} and \eqref{eq:surftens}, respectively, implying that $\mu_{SL}$ is up to sign and units a usual chemical potential.}
Note, however, that the $\mu_{SL}$ defined in (\ref{eq:mu}) will not remain constant when the substrate's bulk elasticity is incorporated.

We thus conclude that shape variations of the solid-liquid interface are governed by the surface tension $\Upsilon_{SL}$, while its composition involves the surface-chemical potential $\mu_{SL}$. This is perfectly in line with previous results derived in Lagrangian formalism \cite{PAKZ2020prx}. 

\subsection{Constitutive relation for the solid interface}
In a previous work \cite{PAKZ2020prx}, we proposed the constitutive relation for the surface elasticity of the solid-liquid interface as
\begin{align}
\label{eq:gammaSL}
\gamma_{SL}(\lambda) &= \gamma_{SL}^0 \left( 1 - c_0 \log \lambda + c_1 (\lambda-1) \right).
\end{align} 
This empirical form reduces to a linear ``surface elasticity" used previously \cite{XuSD2018sm,GKAS2020sm} when expanding around the minimum for small strains. A convenient property of the proposed nonlinear form is that it diverges for $\lambda \to 0$, avoiding a singular mapping. Thermodynamic admissibility requires $\gamma$ to remain positive and convex, which puts constraints on the values of $c_0$ and $c_1$.

In the remainder we will focus on the simplified case where $c_0=c_1$, such that the minimal surface energy is attained for the unstretched state $\lambda=1$. With this, we write (\ref{eq:gammaSL}) as
\begin{align}\label{eq:rheology}
\gamma_{SL}(\lambda) =& \gamma_{SL}^0 + \gamma_{SL}^1 \, g(\lambda), \quad \mathrm{with}\quad g(\lambda)  =\lambda-1 -\log(\lambda)
\end{align} 
where the parameter $\gamma_{SL}^1=\gamma_{SL}^0\,c_0$ governs the strength of the Shuttleworth effect; in the linear description of surface elasticity in \cite{XuSD2018sm}, the coefficient $\gamma_{SL}^1$ is referred to as the modulus of surface elasticity. The corresponding surface tension (\ref{eq:surftens}) reads
\begin{equation}
\Upsilon_{SL}(\lambda) 
= \gamma_{SL}^0 + \gamma_{SL}^1  \left[ 2(\lambda-1) - \log \lambda \right].
\end{equation}
The chemical potential then follows as
\begin{equation}\label{eq:gammamu}
\mu_{SL}(\lambda) = \lambda^2 \frac{\partial\gamma_{SL}}{\partial\lambda}= \gamma_{SL}^1 \lambda(\lambda-1).
\end{equation}
In what follows, the liquid will only cover part of the elastic substrate. Then we will use the same expressions (\ref{eq:gammaSL})-(\ref{eq:gammamu}) derived above for the solid-liquid interface as well for the solid-vapor interface, replacing the subscript ``$SL$" by ``$SV$". Further, we will distinguish the cases of symmetric ($\gamma_{SL}^1=\gamma_{SV}^1$) and asymmetric ($\gamma_{SL}^1\neq\gamma_{SV}^1$) Shuttleworth effect.

\section{Wetting}
\subsection{Macroscopic approach}\label{sec:macro}
In the macroscopic description of wetting, the contact line represents a sharp boundary between the ``wet" and the ``dry" regions of the substrate. On a wet substrate, the solid-liquid interface energy is denoted $\gamma_{SL}(\lambda)$. Similarly, on a dry substrate the solid-vapor energy reads $\gamma_{SV}(\lambda)$, which like $\gamma_{SL}$ will in general be a function of the local stretch. At the contact line the fluid-solid-surface energy is discontinuous in general, and jumps from $\gamma_{SL}$ to $\gamma_{SV}$.

When the liquid is at equilibrium on a rigid homogeneous substrate, the energy of the system must be invariant with respect to a virtual displacement of the contact line along the substrate. Such an equilibrium is only possible when the substrate is perfectly homogeneous, so that the contact line does not exhibit any pinning to a material point on the solid. In this case, energy minimization leads to Young's law for the contact angle, i.e.,
\begin{equation}
\gamma_{LV} \cos \theta_Y =\gamma_{SV} - \gamma_{SL}. 
\end{equation}
On soft substrates, the situation is much more intricate since there are two distinct, independent types of virtual displacements possible at the contact line~\cite{AnSn2020arfm}: (i) Eulerian displacement, exploring the variation of the horizontal and vertical contact line position in the lab-frame, (ii) Lagrangian displacement, exploring the variation of the substrate's material point that is located at the contact line. At equilibrium, where there is no contact line pinning to a specific material point, the energy should be minimal with respect to both kinds of virtual displacements. Variation (i) has been shown to lead to Neumann's law at the contact line~\cite{SnRA2018prl,PAKZ2020prx}. Variation (ii) is needed to prevent pinning to a material point, and gives a second local condition: 
\begin{equation}\label{eq:mueq}
\mu_{SL} = \mu_{SV}.
\end{equation}
This relation expresses that the surface-chemical potential $\mu$ as defined in (\ref{eq:mu}) needs to be continuous across the contact line. 
It was shown that (\ref{eq:mueq}) indeed leads to liquid contact angles, measured with respect to the horizontal, that satisfy Young's law for infinitely large drops -- when drops are large compared to typical elastic deformations \cite{PAKZ2020prx}. However, the equality of chemical potentials across the contact line is a local condition at the contact line, independently of the drop size. Till date, (\ref{eq:mueq}) was only explored for infinitely large drops. Here we will extend this to the case where substrate deformations are comparable to the drop size, for which the liquid angle is known to decrease \cite{StDu2012sm,Lima2012epje,StDuPRL2013,LWBD2014jfm,AnSn2020arfm}; see also Fig.~\ref{fig:sinking}.

\subsection{Mesoscopic approach}
\label{sec:meso}
The  macroscopic features of the contact line, as discussed above,  should emerge naturally in a mesoscopic description, which explicitly accounts for the finite range of molecular interactions. In the mesoscopic framework the transition from the ``wet" to ``dry" is not perfectly sharp, and hence the contact line itself is not sharp. Instead, it becomes a contact-line region described by a continuous function that interpolates between the wet and the dry state. This is achieved by supplementing the surface energy (\ref{eq:fcapbis}) by a wetting energy
\begin{equation}\label{eq:fwet}
\mathcal F_{\rm wet}[h,\xi,u] = \int dx \, f(h,\lambda)\, m(\xi'),
\end{equation}
where we introduce the wetting potential $f(h,\lambda)$, which in principle can depend on the stretch $\lambda$. In the limit where the liquid layer thickness lies outside the range of molecular interactions, one recovers the macroscopic description with a total surface energy as described by (\ref{eq:fcapbis}).  We thus require a wetting potential that on the one hand vanishes as $h\to \infty$. On the other hand, for standard wetting potentials  the ``dry'' substrate corresponds to an adsorption layer of thickness $h_a$, for which $\frac{\partial f}{\partial h}|_{h=h_a}=0$ \cite{Genn1985rmp,TSTJ2018l}. So, for $h=h_a$ the combined effect of $\gamma_{LV} + \gamma_{SL}(\lambda)$ augmented with the wetting potential $f(h_a,\lambda)$ should recover the macroscopic solid-vapor energy, i.e. 
\begin{equation}\label{eq:consistency}
\gamma_{SV}(\lambda) = \gamma_{LV} + \gamma_{SL}(\lambda) + f(h_a,\lambda).
\end{equation}
In consequence, the total mesoscopic capillary energy can be written as the sum of (\ref{eq:fcapbis}) and (\ref{eq:fwet}).
Then, the resulting mesoscopic surface-chemical potential is
\begin{equation}\label{eq:meso-mu}
\mu = \lambda^2 \frac{\partial}{\partial\lambda}\left[\gamma_{SL}(\lambda) + f(h,\lambda)\right].
\end{equation}
Similarly, the mesoscopic surface tension follows as
\begin{equation}\label{eq:meso-upsilon}
\Upsilon =\gamma_{SL}(\lambda) + f(h,\lambda) +  \lambda \frac{\partial}{\partial\lambda}\left[\gamma_{SL}(\lambda) + f(h,\lambda)\right].
\end{equation}
Using Young's law, the correspondence between the mesoscopic and the macroscopic description (\ref{eq:consistency}) can be rewritten as
\begin{equation}\label{eq:Young}
 f(h_a,\lambda) = - \gamma_{LV}\left(1- \cos \theta_Y(\lambda) \right).
\end{equation}
This relates the wetting potential to the macroscopic Young's angle $\theta_Y(\lambda)$, which now depends on $\lambda$. We remind, however, that on elastic substrates Young's law is valid only for drops that are very large as compared to the wetting ridge.

We remark that the energy due to molecular interactions, would in general be a more complex functional that depends on the entire shape of the liquid domain. When the layer is nearly flat, however, the functional reduces to a simple dependence on the local layer thickness, as is assumed above. Strictly speaking, the presented formulation of molecular interactions is thus only valid in the long-wave limit where all interface slopes are small. However, such a mesoscopic model also shows the correct behavior for larger contact angles \cite{HuTA2015jcp}. We will comment on this in more detail when presenting the complete mesoscopic elasto-capillary model.

\subsection{Symmetric vs asymmetric Shuttleworth effect}
We can now distinguish two different scenarios that we will refer to as \emph{symmetric} versus \emph{asymmetric Shuttleworth effect}. In the symmetric case, the wet ($\gamma_{SL}$) and dry ($\gamma_{SV}$) energies exhibit the same dependence on $\lambda$, i.e., in (\ref{eq:rheology}) one has $\gamma^1_{SV}=\gamma^1_{SL}$. Then, identity (\ref{eq:consistency}) conveys that the mesoscopic wetting potential only depends on film thickness, but not on stretch, i.e.\ $f(h,\lambda)=f(h)$. In this case, (\ref{eq:Young}) implies that Young's angle is independent of the stretch. Such a situation was indeed observed in experiments of drops on elastomers where the liquid angle $\theta_ L$, which was assumed $\approx \theta_Y$, was found unaffected when stretching the substrate \cite{STSR2018nc}, even though for some systems a Shuttleworth effect was identified \cite{XJBS2017nc,SnRA2018prl}. 
Furthermore, (\ref{eq:gammamu}) indicates that the functional dependence of $\mu_{SV}(\lambda)$ is the same as that of $\mu_{SL}(\lambda)$. The equality of chemical potential (\ref{eq:mueq}) then amounts to the stretch $\lambda$ being continuous across the contact line. 

In general, however, we need to consider the possibility of an \emph{asymmetric Shuttleworth effect}, macroscopically corresponding to $\partial \gamma_{SV}/\partial \lambda\neq \partial \gamma_{SL}/\partial \lambda$,
i.e., in (\ref{eq:rheology}) one has $\gamma^1_{SV}\neq\gamma^1_{SL}$, and due to Eq.~(\ref{eq:consistency}) the mesoscopic wetting potential depends on stretch as
\begin{equation}
\left.\frac{\partial f}{\partial \lambda}\right|_{h=h_a}
=\frac{\partial \gamma_{SV}}{\partial \lambda} -\frac{\partial \gamma_{SL}}{\partial \lambda}.
\end{equation}
This difference in the strength of the Shuttleworth effect in the wet and dry states renders condition (\ref{eq:mueq}) nontrivial. In this case one expects $\theta_Y$ to depend on the imposed stretch; a stretch-dependent $\theta_L$ as was indeed observed on stiff glassy polymer substrates \cite{STSR2018nc}. Therefore, both the symmetric and asymmetric Shuttleworth effect are of interest. 

\subsection{Specific wetting energy}

While the above expressions are general, we need to make a specific choice for $f(h,\lambda)$ in order to perform calculations. We first recall the specification of the macroscopic surface energies as
\begin{eqnarray}
\gamma_{SL}(\lambda) &=& \gamma_{SL}^0 + \gamma_{SL}^1 \,g(\lambda) \label{eq:gammaslmacro} \\
\gamma_{SV}(\lambda) &=& \gamma_{SV}^0 + \gamma_{SV}^1 \,g(\lambda), \label{eq:gammasvmacro}
\end{eqnarray}
with $g(\lambda)$ already defined in Eq.~\eqref{eq:rheology}. Then, in the mesoscopic description, we propose a product form
\begin{equation}\label{eq:product}
f(h,\lambda) = \nu(\lambda)\,\tilde f(h)
\end{equation}
where the stretch-dependence is encoded via an empirical dimensionless function $\nu(\lambda)$.
The correspondence between the mesoscopic and macroscopic approaches is found via the consistency condition (\ref{eq:Young}), which becomes 
\begin{equation}\label{eq:consistencybis}
\nu(\lambda)\,\tilde f(h_a)   
 = \gamma_{LV}(\cos \theta_Y^0 - 1)  
 + \left( \gamma_{SV}^1 - \gamma_{SL}^1\right) g(\lambda).
\end{equation}
Here we introduced $\theta_Y^0$ as the Young angle at the unstretched state ($\lambda=1$), defined as
\begin{equation}\label{eq:Youngzero}
    \tilde f(h_a)=\gamma^0_{SV} - \gamma^0_{SL}
    - \gamma_{LV} = \gamma_{LV} \left(\cos \theta_Y^0-1 \right).
\end{equation}
We base the thickness-dependent part of the wetting potential on a commonly used, regularized van der Waals interaction for partially wetting liquids on a rigid substrate. In particular, 
\begin{equation}\label{eq:wettPot}
\tilde f(h) = \frac{A}{2h^2}\left[\frac{2}{5}\left(\frac{h_a}{h}\right)^3-1\right].
\end{equation}
where, $A>0$ is the Hamaker constant.
Introducing (\ref{eq:wettPot}) at $h=h_a$ into (\ref{eq:consistency}), we thus require the stretch-dependence to be:
\begin{align}\label{eq:stretchconsist}
\nu(\lambda) = -\frac{10h_a^2}{3A}\left[ \gamma_{LV}(\cos \theta_Y^0 - 1)  
 + \left( \gamma_{SV}^1 - \gamma_{SL}^1\right) g(\lambda) \right].
\end{align}
which finally gives 
\begin{align}\label{eq:fhlambda}
   f(h,\lambda)=\left[\frac{5}{3}\left(\frac{h_a}{h}\right)^2-\frac{2}{3}\left(\frac{h_a}{h}\right)^5\right]\left[\gamma_{LV}(\cos\theta_Y^0-1)+(\gamma_{SV}^1-\gamma_{SL}^1)g(\lambda)\right].
\end{align}
As such, the wetting behavior is  specified by the adsorption  thickness $h_a$, the energies $\gamma_{LV}$, $\gamma_{SL}^0,\gamma_{SV}^0$, and the Shuttleworth coefficients $\gamma_{SL}^1,\gamma_{SV}^1$.

\section{Two elasto-capillary models}\label{sec:twomodels}
The soft wetting problem with Shuttleworth effect can be closed upon introducing the bulk elastic energy of the substrate. Below we propose two different approaches that will be employed, each of which has its own benefits (and drawbacks): 

\begin{itemize}
\item {\bf Macroscopic Neo-Hookean model.} This in principle offers the most complete description of the bulk elasticity of the substrate, resolving the interior stress while consistently accounting for large deformations. This substrate will be coupled to the macroscopic description of wetting. 
\item {\bf Mesoscopic gradient dynamics model.} We use a reduced description of the bulk elasticity by resorting to a Winkler foundation model. When coupled to the mesoscopic description of wetting, this reduced model enables a description of the dynamics of soft wetting. 
\end{itemize}
Below we define both modeling approaches and discuss their numerical implementation. The results from the two approaches will be compared in detail in Sec.~\ref{sec:results}. 
\subsection{Macroscopic Neo-Hookean model}
The Neo-Hookean model for (macroscopic) soft wetting was presented in detail in Pandey \emph{et al.}~\cite{PAKZ2020prx} for deformations induced by a single contact line. Here we extend the formalism to droplets of finite (two-dimensional) volume. A hyperelastic solid is characterized by an energy density $W(\mathbf F)$, where $\mathbf F=\frac{\partial \mathbf r}{\partial \mathbf R}$ is the (gradient) deformation tensor. In two dimensions, the combined elastic and capillary energy (per unit length) reads
\begin{eqnarray}
\label{eq:Echi}
\mathcal F[\chi] 
= \int dXdY \, W(\mathbf F) 
+ \int dX \, \lambda \gamma(\lambda),
\end{eqnarray}
where $\gamma$ may stand for $\gamma_{SL}$ or $\gamma_{SV}$, depending on whether the surface is locally wet or dry. 
This energy is a functional of the mapping $\mathbf r = \chi(\mathbf R)$. Since the hyperelastic description is Lagrangian, we have also expressed the surface energy as an integral over $X$ at $Y=0$. To account for the correct surface metric, we used the connection  $ds=\lambda dX$, where $\lambda$ is the stretch at the surface [cf.~(\ref{eq:stretch})]. In the calculations below we use an incompressible Neo-Hookean energy density, which in two dimensions reads
\begin{equation}
W(\mathbf F) = \frac{1}{2}G \left( {\rm tr}(\mathbf F \cdot \mathbf F^T) - 2\right) - p \left(\mathrm{det}(\mathbf F)- 1 \right).
\end{equation} 
Here $G$ is the shear modulus, while we have included the constraint of incompressibility via the Lagrange multiplier $p$.

\begin{figure}[H]
\centerline{\includegraphics[width=\textwidth]{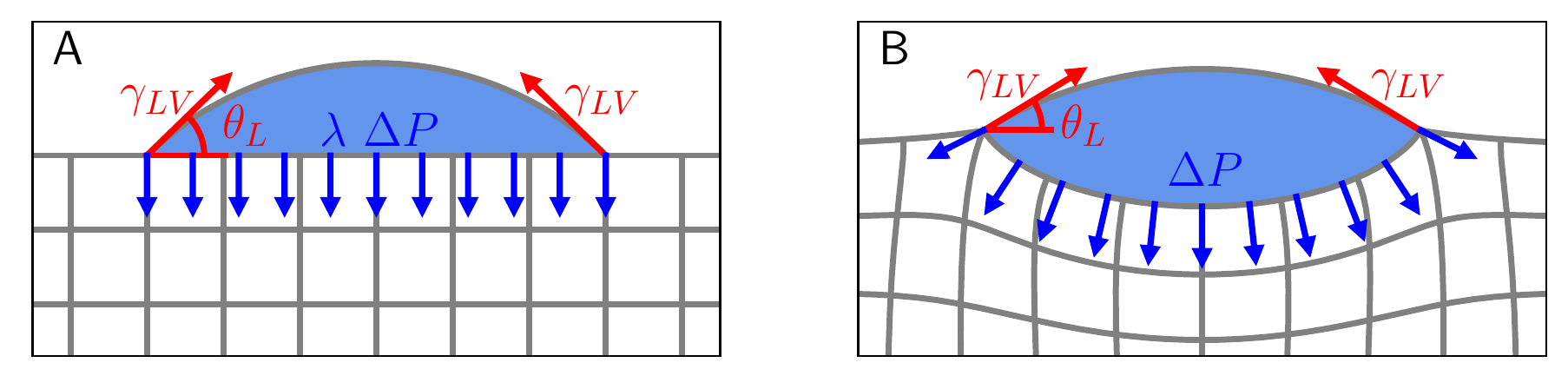}}
\caption{The traction exerted by the droplet onto the substrate, consisting of two localized contact line forces and the Laplace pressure $\Delta P$ inside the drop. The liquid contact angle $\theta_L$, measured with respect to the horizontal in both the (a) reference and (b) deformed configurations, is not known a priori, but determined consistently from (\ref{eq:mueq}).}
\label{fig:traction}
\end{figure}

The wetting is accounted for via the traction that is exerted by the drop onto the substrate. This traction is sketched in Fig.~\ref{fig:traction}. It consists of two localized forces $\gamma_{LV}\mathbf t$ pulling along the liquid-vapor interface at the two contact lines, located at $X=\pm R$ and $Y=0$. Here $\mathbf t$ is the tangential unit vector, i.e., the force pulls at an angle $\theta_L$. 
It is noteworthy that the localized loads would lead to an ill-posed minimization problem in the absence of solid surface energy, and that the solid surface energy provides sufficient regularization to render the minimization problem well-posed.
In between the contact lines the droplet's (Laplace) pressure $\Delta P$ is exerted on the substrate. It is related to the liquid angle $\theta_L$ as $\Delta P = \gamma_{LV} \sin\theta_L/r$, with $r$ being the (Eulerian) base radius of the droplet. Formally, this traction is captured by a work functional
\begin{equation}\label{eq:work}
\mathcal R[\chi] = \gamma_{LV} \mathbf t_{R} \cdot \mathbf r(R,0) + \gamma_{LV}  \mathbf t_{-R} \cdot \mathbf r(-R,0)   - \int_{-R}^R dX \, \lambda  \Delta P \, \mathbf n \cdot \mathbf r(X,0),
\end{equation}
where $\mathbf n$ is the surface normal in the current configuration. The problem is then defined by minimization of $\mathcal F - \mathcal R$, with respect to the mapping $(X,Y) \mapsto \mathbf r =\chi(X,Y)$. Importantly, the Neumann condition at the contact line emerges within this framework, since the minimization is explicitly done with respect to the Eulerian contact line position, $\delta \mathbf r$. However, the work functional (\ref{eq:work}) still contains an unknown liquid angle $\theta_L$; this angle can be found by imposing the no-pinning condition (\ref{eq:mueq}), which reflects the variation of the Lagrangian contact line position (see the discussion in Sec.~\ref{sec:macro}). The problem is therefore closed by introducing the liquid angle as an additional variable, with the no-pinning condition (\ref{eq:mueq}) as the corresponding residual. 

In summary, the elastocapillary problem thus consist of minimizing the functional
\begin{align}
\mathcal F[\chi] - \mathcal R[\chi] = 
&\iint dXdY \, W(\mathbf F) 
+\int dX \, \lambda \gamma(\lambda)\\
-&
 \gamma_{LV} \mathbf t_{R} \cdot \mathbf r(R,0) - \gamma_{LV}  \mathbf t_{-R} \cdot \mathbf r(-R,0)
+\int_{-R}^R dX \, \lambda  \Delta P \, \mathbf n \cdot \mathbf r(X,0),
\nonumber
\end{align}
subject to the no-pinning condition $\mu_{SV}=\mu_{SL}$ at the contact line, to consistently determine the equilibrium liquid angle $\theta_L$. The minimization of the energy functional $\mathcal F - \mathcal R$ is based on the method in \cite{PAKZ2020prx}, adapted to the specific problem at hand. For simplicity the \emph{goal-adaptive} finite-element method used in \cite{PAKZ2020prx} is replaced by a \emph{residual-based} method, in which elements are selected for refinement based on the residuals when the current solution is projected on a refined mesh. This method is implemented using the open-source numerical framework Nutils \cite{ZZVF2020nutils}.

The solid substrate measures $8R \times 8R$ in the undeformed configuration. The left and right boundaries of the substrate are only fixed in horizontal direction,  allowing for movement in the vertical direction. The bottom boundary is fixed in both directions. We verified that results are nearly independent of domain-size. For example, doubling the thickness from 8R to 16R, the angle changes by $5\times 10^{-5}$ degrees and the wetting ridge height by $1\times 10^{-6}$.
The substrate is initially divided into a mesh of $48 \times 48$ elements and subsequently undergoes a total of $13$ refinement iterations. At maximum refinement the element size is reduced by a factor $2^{-13}$, and a minimum element size of approximately $2R\times10^{-5}$ is reached. Since these elements are significantly smaller than the elastocapillary length, this ensures that wetting ridges are accurately resolved.

\subsection{Mesoscopic gradient dynamics model}
\label{sec:mesomodel-summ}
The second approach is in the spirit of the gradient dynamics approach (see, e.g., \cite{ThAP2016prf,Thie2018csa}) to the dynamics of drops on simple compressible elastic substrates presented by Henkel \emph{et al.}~\cite{HeST2021sm}, using a mesoscopic wetting description as given in section~\ref{sec:meso}. In this approach the hyperelastic model for the bulk elasticity is replaced by a simpler ``Winkler-type'' approximation, for which the elastic energy depends only on the displacements of the interface. Using this reduced elastic energy together with a compressible substrate dynamics coupled to a mesoscopic model for the dynamics of the liquid (thin-film, long-wave or lubrication model \cite{OrDB1997rmp,Thie2010jpcm}) one obtains a versatile modeling framework. In contrast to the hyperelastic model it allows one to study dynamical effects like viscoelastic braking in droplet spreading as well as film dewetting and subsequent coarsening of ensembles of drops on elastic substrates \cite{HeST2021sm}. Here we extend this type of mesoscopic model to incorporate the Shuttleworth effect considering full-curvature \cite{Snoe2006pf,Thie2018csa} and long-wave versions.

The total free energy of the gradient dynamics model is a functional of the scalar Eulerian fields $h(x,t),\xi(x,t),u(x,t)$, and reads
\begin{equation}\label{eq:thinfilm}
\mathcal F[h,\xi,u] = \mathcal F_{\rm el}[\xi,u] + \mathcal F_{\rm cap}[h,\xi,u] + \mathcal F_{\rm wet} [h,\xi,u],
\end{equation}
with the capillary and wetting energies defined above, respectively, in (\ref{eq:fcapbis}) and (\ref{eq:fwet}). The elastic energy is approximated by 
\begin{equation}\label{eq:winkler}
\mathcal F_{\rm el}[\xi,u] = \int dx \, \frac{\kappa}{2} \left( \xi^2 + u^2 \right),
\end{equation}
which involves an integral only over the interface (and not over the substrate depth, as is the case for the Neo-Hookean model). The Winkler foundation model employed in \cite{HeST2021sm} only describes the vertical displacement $\xi$, where $\kappa$ is the effective stiffness of the substrate. In \eqref{eq:winkler} we have now added a rigidity with respect to lateral displacements. For reasons of simplicity we use the same effective stiffness $\kappa$.
To enable the possibility of a prestretched substrate (as is common in experiments and in the Neo-Hookean model), we adapt the energy as
\begin{equation}\label{eq:winklerstretch}
\mathcal F_{\rm el}[\xi,u] = \int dx \, \frac{\kappa}{2} \left[ \xi^2 + \left( u - u'_\infty x \right)^2  \right],
\end{equation}
where $u'_\infty$ corresponds to an imposed prestretch  $\lambda_\infty=1/(1-u'_\infty)$  prior to placing a droplet. 

The static, equilibrium properties of a drop of some finite volume $V$ can be inferred by minimizing (\ref{eq:thinfilm}) together with the condition for volume conservation $\int h dx =V$ with respect to the three steady fields $h(x),\xi(x)$ and $u(x)$. However, the formulation furthermore naturally admits a gradient dynamics structure that, as a bonus, gives a time evolution towards this equilibrium. For this we consider the time-dependent fields $h(x,t),\xi(x,t)$ and $u(x,t)$, and define the gradient dynamics model
\begin{eqnarray}
\frac{\partial h}{\partial t} &=& \frac{\partial}{\partial x}\left[ \frac{h^3}{3\eta} \frac{\partial}{\partial x} \left(\frac{\delta \mathcal F}{\delta h}\right) \right] \label{eq:graddyn1}\\
\frac{\partial \xi}{\partial t} &=& - \frac{1}{\zeta} \frac{\delta \mathcal F}{\delta \xi} \label{eq:graddyn2} \\
\frac{\partial u}{\partial t} &=& - \frac{1}{\zeta} \frac{\delta \mathcal F}{\delta u}. \label{eq:graddyn3}
\end{eqnarray}
where we assumed the same ‘‘elastic friction constant’’ $\zeta$ governs the relaxation of $\xi$ and $u$. As for the considered nonabsorbing substrate there is no mass transfer between the liquid layer and the elastic substrate and the considered liquid is nonvolatile, the liquid dynamics (\ref{eq:graddyn1}) is fully conserved. The nonconserved dynamics (\ref{eq:graddyn2}) and (\ref{eq:graddyn3}) for the deformations $\xi$ and $u$, respectively, reflect the assumed full compressibility of the elastic substrate.
For the derivation of such equations based on the Onsager variational principle see, e.g., \cite{Doi2011jpcm,Thie2018csa}.

The variations of \eqref{eq:winkler} are
\begin{eqnarray}\label{eq:varimeso-h}
\frac{\delta F}{\delta h} &=& 
-\gamma_{LV}\frac{\partial}{\partial x}\left(\frac{h'+\xi'}{m(h'+\xi')}\right) +\frac{\partial f}{\partial h}m(\xi')\\
\frac{\delta F}{\delta \xi} &=& -\gamma_{LV}\frac{\partial}{\partial x}\left(\frac{h'+\xi'}{m(h'+\xi')}\right) -\frac{\partial}{\partial x}\Upsilon\left(\frac{\xi'}{m(\xi')}\right)+\kappa\,\xi \label{eq:varimeso-xi}\\
\frac{\delta F}{\delta u} &=& -\frac{\partial \mu}{\partial x} + \kappa\,(u-u'_\infty x)
\label{eq:varimeso-u}
\end{eqnarray}
where $m(z)=\sqrt{1+z^2}$ is again the metric factor. The variation with respect to $h$ expresses the (liquid-vapor) capillary pressure and the disjoining pressure due to the molecular interactions. The variation with respect to $\xi$ expresses the capillary pressures and the substrate elasticity. Finally, the variation with respect to $u$ controls the substrate's composition, leading to a shift of $\mu$ due to elasticity. In the long-wave approximation (valid at small slopes) the above expressions can be simplified (see Appendix~\ref{app-longwave}).
Other dynamic long-wave models without considering the Shuttleworth effect or lateral displacements were developed for the dynamics of a liquid drop on a viscoelastic layer \cite{MaGK2005jcis,GLBG2017csaea,ChKu2020sm} and for the durotaxis of a liquid drop on a compliant Kirchhoff plate \cite{GoVe2020epjst} while certain elasticity aspects also enter long-wave models for drops on polymer brushes \cite{ThHa2020epjt} and on growing layers of ice  \cite{SLNA2021nc}. These long-wave descriptions are further discussed in section~2.1 of Ref.~\cite{HeST2021sm}.

Eqs.~(\ref{eq:graddyn1})-(\ref{eq:graddyn3}) as well as their long-wave equivalents (Appendix~\ref{app-longwave}) are simulated in time employing the FEM-based software package \textsc{oomph-lib} \cite{HeHa2006}. An adaptive time stepping is used based on a backward differentiation method of order 2 (BDF2) from which the next state is obtained via a Newton procedure. The efficient adaptive time stepping and mesh refinement routines allow for a treatment of even very large systems. Branches of steady states are as well followed in parameter space employing the continuation routines \cite{DWCD2014ccp,EGUW2019springer,Thiele2021lectureCont} bundled in \textsc{pde2path} \cite{UeWR2014nmma}.

Finally, note a peculiar property of the chosen elasticity model and setting without additional body forces: Even though the elastic layer is locally compressible, all steady states (characterized by $\delta F/\delta \xi=0$) have a zero global vertical displacement $\Xi=\int\xi dx=0$ (when using periodic or Neumann boundary conditions). This is seen when integrating \eqref{eq:varimeso-xi} over the domain. When similarly integrated, the nonconserved dynamics \eqref{eq:graddyn2} reduces to $\partial \Xi / \partial t = - (\kappa/\zeta)\,\Xi$, i.e., $\Xi=0$ is a stable fixed point. The described behavior directly follows from the simple parabolic elastic energy \eqref{eq:winkler}, i.e., the Winkler foundation model. The inclusion of a body force like gravity shifts this fixed point away from zero. For comparison, the incompressible  neo-Hookean substrate is  strictly volume conserving, locally and globally, also in the presence of body forces.

\subsection{Model parameters and the elastocapillary length}
The two models contain various different parameters, so great care must be taken when comparing the results. The parameters are summarized in Table~\ref{tab:parameters2}. The macroscopic surface energies can be chosen identical in both models, and require a choice for 
 the energy coefficients  $\gamma_{LV}$, $\gamma_{SL}^0,\gamma_{SV}^0$, and the Shuttleworth coefficients  $\gamma_{SL}^1,\gamma_{SV}^1$, as defined in (\ref{eq:gammaslmacro}) and (\ref{eq:gammasvmacro}). The mesoscopic model contains the adsorption layer thickness $h_a$ as an additional parameter. We choose $h_a$ to be sufficiently small such that it does not affect the macroscopic elastic deformations and the contact angle of the drop. 

\begin{table}[H]
\caption{Summary of parameters in the macroscopic and mesoscopic models. The connection of macroscale parameters $\gamma_{SV}^0$, $\gamma_{SV}^1$ and mesoscale wetting potential $f(h,\lambda)$ is given by the consistency conditions (\ref{eq:Youngzero}) and \eqref{eq:stretchconsist}.}
\label{tab:parameters2}
\centering
	\begin{tabularx}{0.65\textwidth}{l|c|c}
		\hline
         \textbf{Quantity} & \textbf{Macroscopic} & \textbf{Mesoscopic}  \\
         \hline
         Surface energies ($\lambda=1$)& $\gamma_{LV},\gamma_{SL}^0,\gamma_{SV}^0$ & $\gamma_{LV},\gamma_{SL}^0$  \\
          Shuttleworth constants
          &$\gamma_{SL}^1,\gamma_{SV}^1$ &$\gamma_{SL}^1$  \\
          Adsorption layer thickness & - & $h_a$ \\
          Wetting potential & - & $f(h,\lambda)=\nu(\lambda)\,\tilde f(h)$  \\
          \hline
         Substrate stiffness & $G$ &$\kappa$\\
         Elasto-capillary length& $\gamma_{LV}/G$ &$\sqrt{\gamma_{LV}/\kappa}$\\
         \hline
         Liquid viscosity & - &$\eta_L$\\
         Elastic friction constant & - &$\zeta$\\
         \hline
	\end{tabularx} 
\end{table}

While the capillarity and wetting energies of the two models can be set to fully agree in the macroscopic limit, this is not the case for the elastic energy. The elasticity of the (incompressible) Neo-Hookean model is described by the shear modulus $G$. In the gradient dynamics model, elasticity is implemented through a Winkler foundation model, which contains an empirical elastic constant $\kappa$. For \emph{compressible} layers, the constant $\kappa$ can be expressed in terms of $G$ using a long-wave expansion for a thin elastic layer \cite{HeST2021sm}. However, the expansion for \emph{incompressible} elastic layers does not reduce to the Winkler form, and the systematic connection cannot be established. However, motivated by Ref.~\cite{HeST2021sm}, the connection between the two models can be made via the elastocapillary length. For the two models it is respectively defined as 
\begin{alignat}{3}\label{eq:elasto_capillary_length}
&\ell_{\rm ec}^{\rm NH} &&= \frac{\gamma_{LV}}{G} && \quad \mathrm{Neo-Hookean}, \\
&\ell_{\rm ec}^{\rm GD} &&= \left(\frac{\gamma_{LV}}{\kappa}\right)^{1/2} && \quad \mathrm{Gradient \, dynamics}.
\end{alignat}
In what follows we will therefore quantify the ``softness'' using $\ell_{\rm ec}/r$. This dimensionless number scales the elastocapillary length of the models to the half-width of the drop (quantified by the contact line position $x=r$ in the deformed configuration). This enables a one-to-one comparison between equilibrium shapes (drop and substrate) obtained in the two models, without any adjustable parameters. 

Besides these energetic parameters, the gradient dynamics model involves dynamical parameters: the viscosity of the liquid layer $\eta$ and the elastic friction constant $\zeta$ that encodes the timescale of the substrate.

\section{Contact angles and substrate deformations}\label{sec:results}

Typical results of the two models are shown in Fig.~\ref{fig:sinking}. On relatively stiff substrates, the droplet induces small wetting ridges at the contact line. Upon decreasing the substrate stiffness the drops gradually sink into the substrate, until attaining a liquid-like geometry. This rigid-to-soft transition is characterized in quantitative detail below, focusing on the liquid contact angle $\theta_L$ and the deformations of the substrate.

\subsection{Contact angles}
\begin{figure}[H]
\centerline{\includegraphics[width=\textwidth]{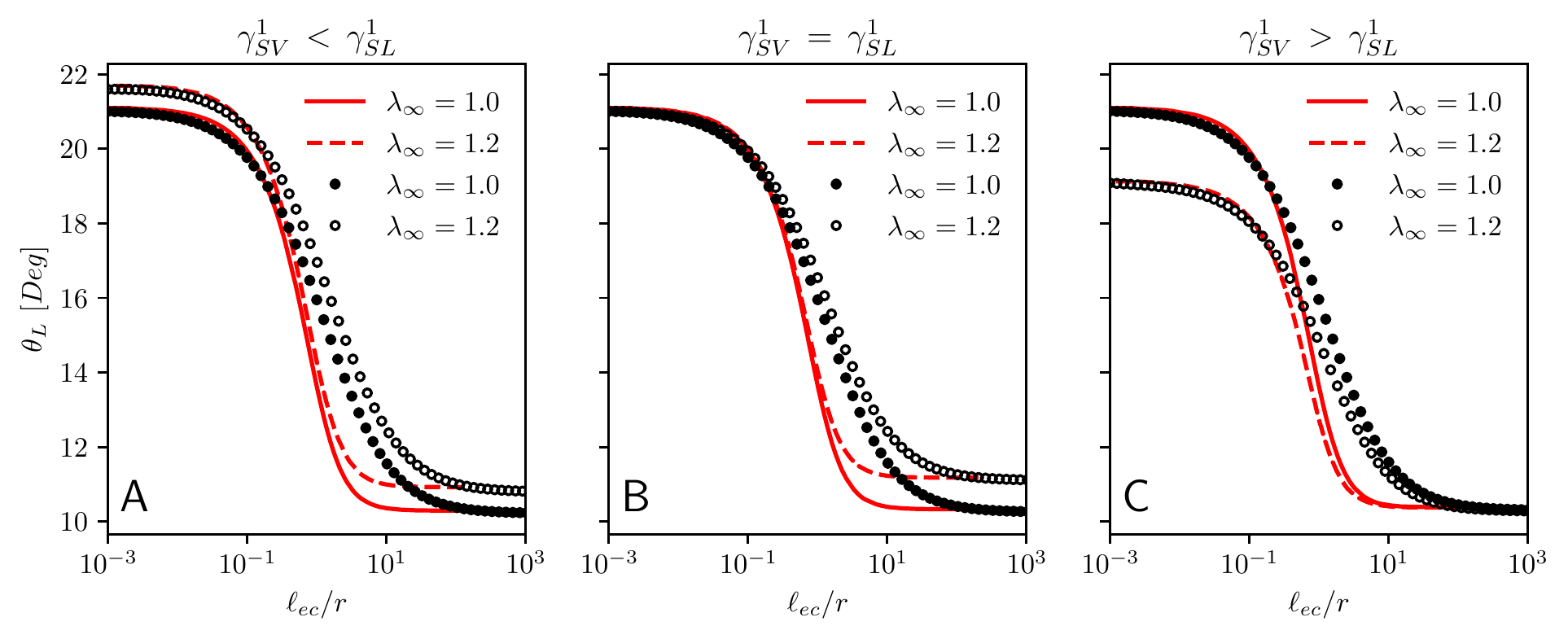}}
\caption{Liquid contact angle $\theta_L$ versus substrate softness $\ell_{\rm ec}/r$, for symmetric and asymmetric Shuttleworth effect. (a) $\gamma^1_{SV} < \gamma^1_{SL}$, (b) $\gamma^1_{SV} = \gamma^1_{SL}$, (c) $\gamma^1_{SV} > \gamma^1_{SL}$. Black symbols and red lines correspond to the macroscopic neo-Hookean and to the mesoscopic gradient dynamics model, respectively.  Results without ($\lambda_\infty=1$) and with ($\lambda=1.2$) prestretch are shown as closed symbols / solid lines and open symbols / dashed lines, respectively. 
Parameter values are $\theta_Y^0=21.06^\circ$, $\gamma_{SL}^{1}=\gamma_{SL}^0$, while $\gamma^1_{SV}$ is chosen $\tfrac{1}{3}$, $1$, or $3$ times $\gamma_{SL}^1$.
}
\label{fig:transition}
\end{figure}

In Fig.~\ref{fig:transition} we report the transition of the liquid contact angle $\theta_L$ between the limiting cases of rigid and liquid substrates as a function of the softness $\ell_{\rm ec}/r$, in the presence of the Shuttleworth effect with ($\lambda_\infty\neq1$) and without ($\lambda_\infty=1$) prestretch. The black symbols correspond to the results of the macroscopic neo-Hookean model, while the red lines represent the mesoscopic gradient dynamics model. In all calculations the surface energies without stretch ($\gamma_{LV}^0,\gamma_{SV}^0,\gamma_{SL}^0$) were fixed to constant values, such that the corresponding Young's angle $\theta_Y^0=21.06^\circ$. All curves exhibit a transition from ``Young'' to ``Neumann'', namely, $\theta_L$ decreases as the substrate gets softer, i.e., as one increases $\ell_{\rm ec}/r$. The details of this transition depend on the choice of the Shuttleworth coefficients $\gamma^1_{SV},\gamma^1_{SL}$ (different panels), and on the prestretch of the substrate ($\lambda_\infty=1$ vs.\ $\lambda_\infty=1.2$, see legends).

\subsubsection{Symmetric Shuttleworth effect}

 Figure~\ref{fig:transition}(b) corresponds to a situation with a symmetric Shuttleworth effect, for which $\gamma^1_{SV}=\gamma^1_{SL}$. In the limit of rigid substrates ($\ell_{\rm ec}/r \ll 1$), we find that the liquid angle is independent of prestretch $\lambda_\infty$. This independence reflects that for a symmetric Shuttleworth effect the difference in surface energies  $\gamma_{SV}-\gamma_{SL}$ is not affected by the imposed $\lambda_\infty$. In other words, stretching a very rigid substrate does not render it more hydrophilic or more hydrophobic. However, the effect of stretching becomes apparent when the substrate is deformable. In the soft limit ($\ell_{\rm ec}/r \gg 1$) of this particular example we find $\theta_L=10.3^\circ$ without prestretch and $\theta_L=11.1^\circ$ for $\lambda_\infty = 1.2$.  This difference in contact angles can be attributed to the changes in the surface tensions due to stretching, which affect the vectorial Neumann's balance (even though Young's angle based on surface energies remains unaffected).

Let us now discuss the predictions by the macroscopic neo-Hookean model (symbols) in comparison to those of the mesoscopic gradient dynamics model (lines). First, we note that both models predict the same angles $\theta_L$ in the rigid and soft limits. This reflects that these limiting values for the liquid angle are solely dictated by capillarity (Young and Neumann, respectively) -- and capillarity is rigorously implemented in both models. However, it is clear that the rigid-to-soft transition is much more abrupt in the gradient dynamics model as compared to the neo-Hookean simulations. The contact angle in the gradient dynamics model sharply changes within about one order of magnitude around $\ell_{\rm ec}/r \sim 1$, while the neo-Hookean model takes two to three orders of magnitude in softness to effectuate the transition. In consequence, the neo-Hookean liquid angles are larger than those in the gradient dynamics model during the transition. We attribute the slow transition for the neo-Hookean solid to the long-range nature of elastic interactions \cite{Johnson1987}: the displacement induced by a localized traction exerted onto a two-dimensional elastic medium decays only logarithmically with distance, until the size of the system is encountered. This long-ranged nature of elasticity is lost when approximating the substrate by Winkler's foundation, for which the relation between traction and displacement is perfectly local. We return to this long-range interaction below, when discussing the substrate deformations.

We thus conclude that the mesoscopic gradient dynamics model with a reduced description of elasticity faithfully reproduces the  equilibrium angles in the rigid and soft limits, including the effect of prestretch. When expressing the stiffness through $\ell_{\rm ec}/r$, the reduced model captures the trends qualitatively, but significant quantitative differences appear in the transition range. Similar observations regarding the two models apply to all panels in Fig.~\ref{fig:transition}.

\subsubsection{Asymmetric Shuttleworth effect}
We now turn to the case of an asymmetric Shuttleworth effect, for which $\gamma^1_{SV}\neq \gamma^1_{SL}$. Figure~\ref{fig:transition}(a) corresponds to a situation with $\gamma^1_{SV}<\gamma^1_{SL}$, such that the solid-liquid energy increases more with stretch than the solid-vapor energy. In this case, the substrate becomes more ``hydrophobic'' once it is stretched. Indeed, one observes larger contact angles $\theta_L$ for $\lambda_\infty=1.2$ as compared to the unstretched case $\lambda_\infty=1$. We verified that in the rigid limit, the increase of $\theta_L$ exactly matches that predicted by Young's law  based on the energies at $\lambda_\infty$. This enhanced $\theta_L$ with stretch is apparent irrespective of the substrate softness.  

The asymmetric Shuttleworth effect with $\gamma^1_{SV} > \gamma^1_{SL}$ is shown in  Fig~\ref{fig:transition}(c). This case is opposite to that of panel (a), since now the substrate becomes more ``hydrophilic'' when stretched. In the rigid limit ($\ell_{\rm ec}/r \ll 1$) one indeed observes smaller contact angles $\theta_L$ for $\lambda_\infty=1.2$ as compared to $\lambda_\infty=1$. Again, this is in accordance with Young's law based on the imposed $\lambda_\infty$. Interestingly, the difference in contact angle is no longer apparent in the soft limit ($\ell_{\rm ec}/r \gg 1$). To predict the contact angle in this soft, Neumann limit, however, is not straightforward: Neumann's balance depends on the local values of surface tensions at the contact line. These local surface tensions depend not on $\lambda_\infty$ but on the local values of the stretches at the contact line, which, as we see below, take on nontrivial values. 

\subsection{Substrate deformations}
\begin{figure}[H]
\centerline{\includegraphics[width=\textwidth]{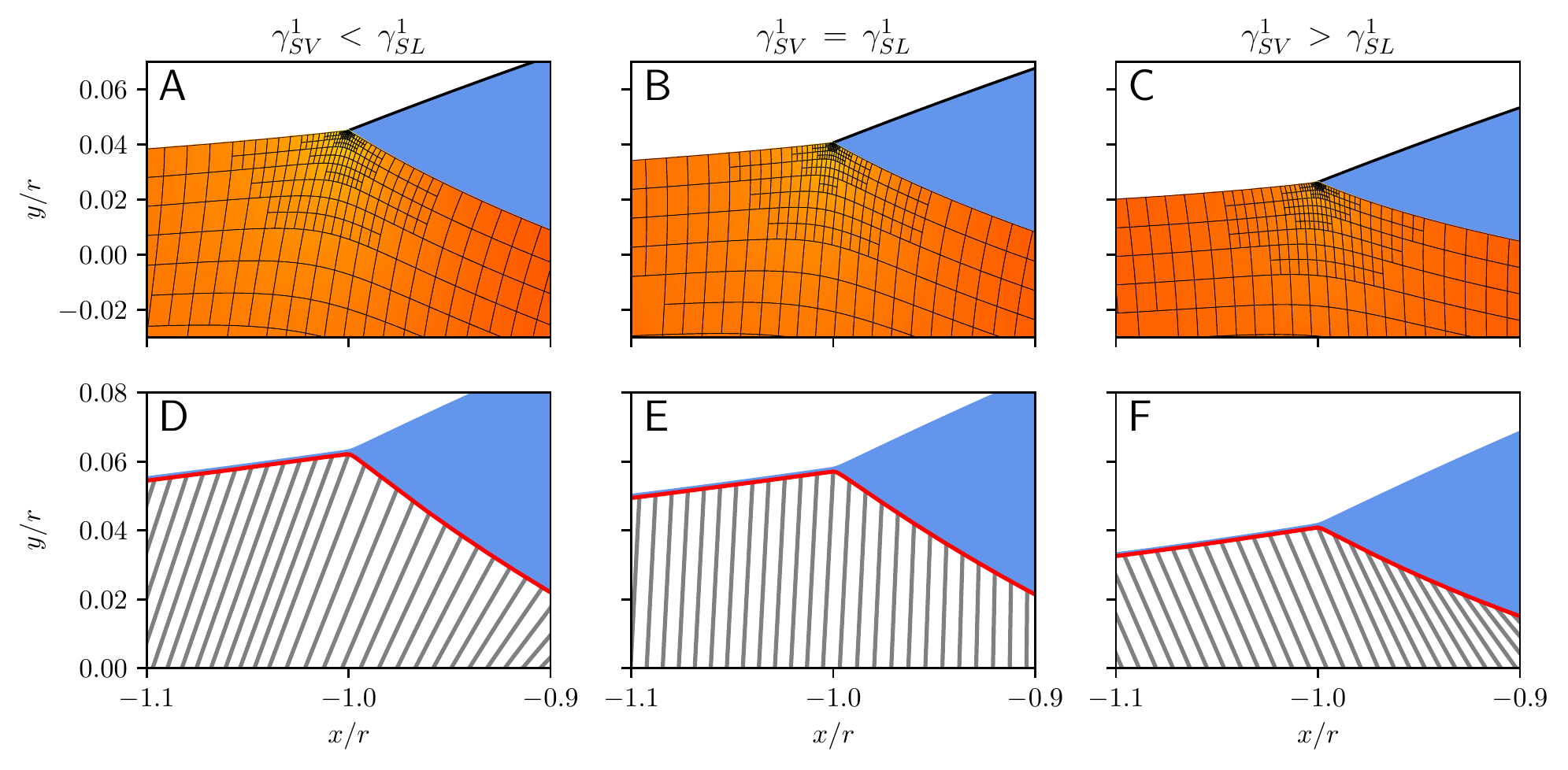}}
\caption{Typical height profiles and substrate deformations in the vicinity of a contact line as obtained with the (a-c) macroscopic and (d-f) mesoscopic model in the cases of symmetric and asymmetric Shuttleworth effect as indicated above the panels.
The central result is that horizontal displacements induced by the droplet are governed by the (a)symmetry of the Shuttleworth effect. In the top row (a-c) the deformation is visible from the grids that in the reference state are straight horizontal/vertical. In the bottom row (d-f) the lines indicate the interface  displacement induced by the presence of the drop.
 The substrate is prestretched with $\lambda_\infty=1.2$. Further parameters are $\theta^0_Y=21.06^\circ$, $\theta_L \approx 16^\circ$, $\gamma^1_{SL} = \gamma^0_{SL}$, and $\gamma^1_{SV}=\tfrac{1}{3}\,\gamma^1_{SL}$ (left) $\gamma^1_{SV}=\gamma^1_{SL}$, (center) $\gamma^1_{SV}=3\,\gamma^1_{SL}$, (right). Note that for each panel $\ell_\mathrm{ec}/r$ is selected by $\theta_L$, cf.~Fig.~\ref{fig:transition}.
}
\label{fig:tangential}
\end{figure}

We now turn to a detailed discussion of the substrate deformations, where once again we investigate the effect of the (a)symmetric Shuttleworth effect in both the macroscopic and the mesoscopic models. 
Figure~\ref{fig:tangential} shows magnifications of the vicinity of the contact line, as obtained within the two models. We selected simulations from Fig.~\ref{fig:transition} with liquid angles $\theta_L \approx 16^\circ$. When comparing the various panels with different Shuttleworth effect, one notices a clear difference in horizontal displacements. 

\begin{figure}[H]
\centerline{\includegraphics[width=\textwidth]{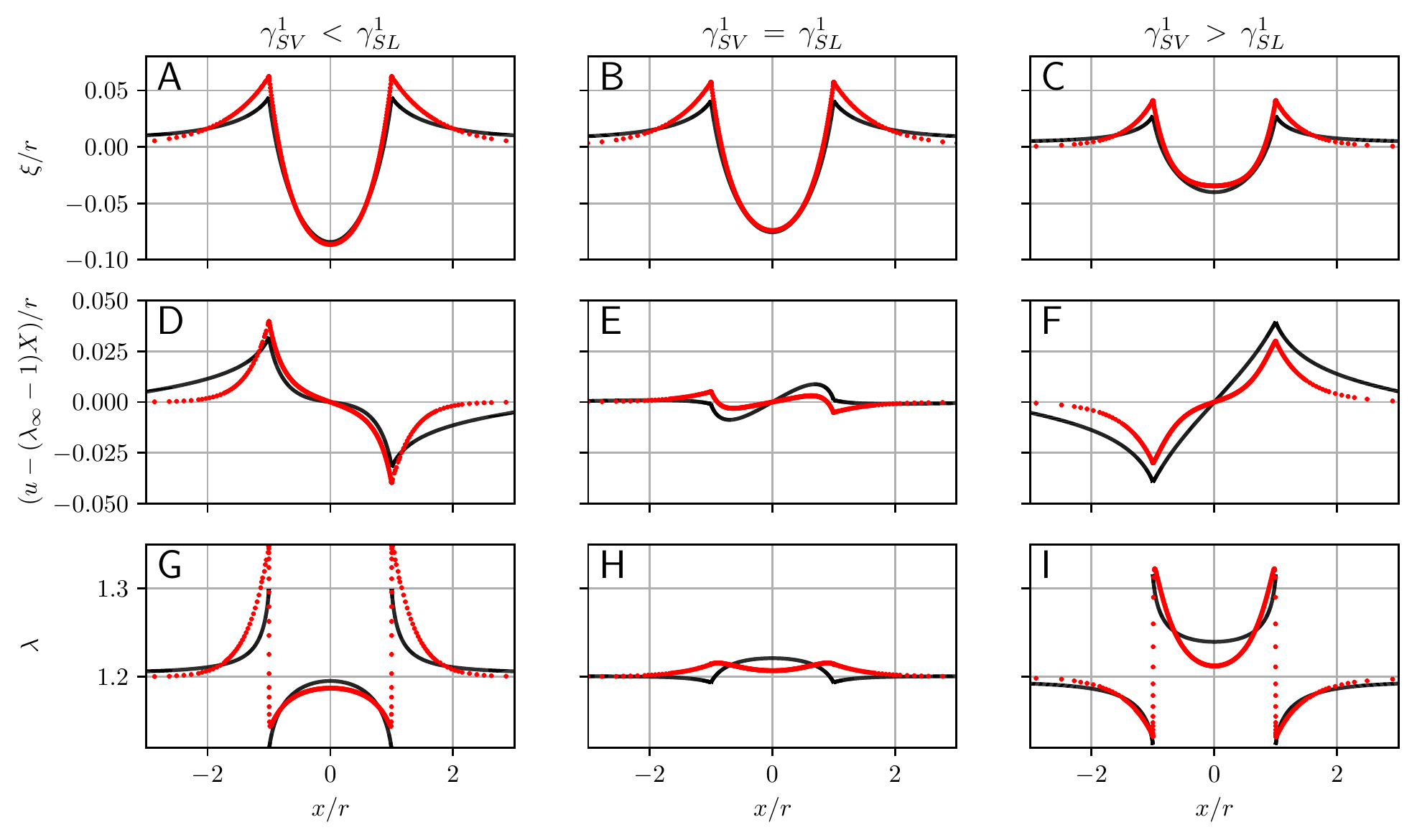}}
\caption{Substrate deformations for symmetric and asymmetric Shuttleworth effect. (top row) Vertical displacements $\xi(x)$, normalized by drop size $r$. (middle row) Horizontal displacements relative to the prestretch, i.e., $u(x)-(\lambda_\infty-1) X$, normalized by drop size $r$. (bottom row) Surface stretch $\lambda(x)$. Black lines correspond to the macroscopic neo-Hookean model, red symbols to the mesoscopic model. Parameters are as in Fig.~\ref{fig:tangential}.
}
\label{fig:deformations}
\end{figure}

Detailed quantitative comparisons are presented in Fig.~\ref{fig:deformations}, where black and red data are obtained with the macroscopic and the mesoscopic model, respectively. To enable a ``fair'' comparison between the two models, we select data at nearly identical liquid angles, at $\theta_L \approx 16^\circ$, which lies halfway the rigid-to-soft transition. The data in Fig.~\ref{fig:deformations} are taken for a prestretch of $\lambda=1.2$.

The top row of Fig.~\ref{fig:deformations} shows the vertical substrate displacements $h(x)$, normalized by the drop size, for symmetric and asymmetric Shuttleworth effect. The profiles all look very similar, with a very good agreement between the neo-Hookean (black) and mesoscopic (red) models. We observe the latter to produce slightly higher ridges than the former model. Away from the drop, the black wetting ridges systematically decay more slowly than the red ones. This signals the previously mentioned long-ranged elastic interactions, which are not faithfully captured by the Winkler foundation used in the mesoscopic model. 

The middle row of Fig.~\ref{fig:deformations} shows the horizontal substrate displacements induced by the droplet, $u(x) - \lambda_\infty X$, where we corrected for the imposed prestretch. Now significant differences appear between the (a)symmetric cases. Comparing the leftmost panel ($\gamma_{SV}^1 < \gamma^1_{SL}$) to the rightmost panel ($\gamma_{SV}^1 > \gamma^1_{SL}$), we observe a change from ``inward'' to ``outward'' horizontal displacements. This can be interpreted along the lines of~Refs.~\cite{WeAS2013sm,AnSn2016el}, who show that -- in the rigid limit -- a resultant horizontal force $\gamma^1_{SL} - \gamma^1_{SV}$ is exerted onto the substrate, oriented towards the droplet. Indeed, here we find that the horizontal displacement changes its orientation when this difference changes sign. Consistently, for the case of symmetric Shuttleworth effect only a very small horizontal displacement is observed. 

Finally, the bottom row of Fig.~\ref{fig:deformations} shows the stretches $\lambda(x)$ along the surface of the substrate. These stretches are subject to the conditions of continuous chemical potential $\mu_{SV}=\mu_{SL}$ across the contact line. In case of symmetric Shuttleworth effect, this continuity of $\mu$ implies a continuity of stretch $\lambda$. Indeed, the middle panel exhibits continuous $\lambda$ at the contact line, with only mild variations around the imposed value of $\lambda_\infty=1.2$. This is in stark contrast to the case of the asymmetric Shuttleworth effect (left and right panels), for which the stretch is observed to exhibit a jump across the contact line -- in the macroscopic model it is truly a discontinuity, while in the mesoscopic model the jump is smooth on the scale of molecular interactions. The jump in stretch is necessary to ensure continuous $\mu$. Overall, stronger variations in $\lambda$ are observed for asymmetric Shuttleworth effect. Thereby the larger $\lambda$ are observed for the interface with the smaller $\gamma^1$, i.e., outside the drop in Fig.~\ref{fig:deformations}(g) and inside the drop in Fig.~\ref{fig:deformations}(i).

\begin{figure}[H]
\centerline{\includegraphics[width=\textwidth]{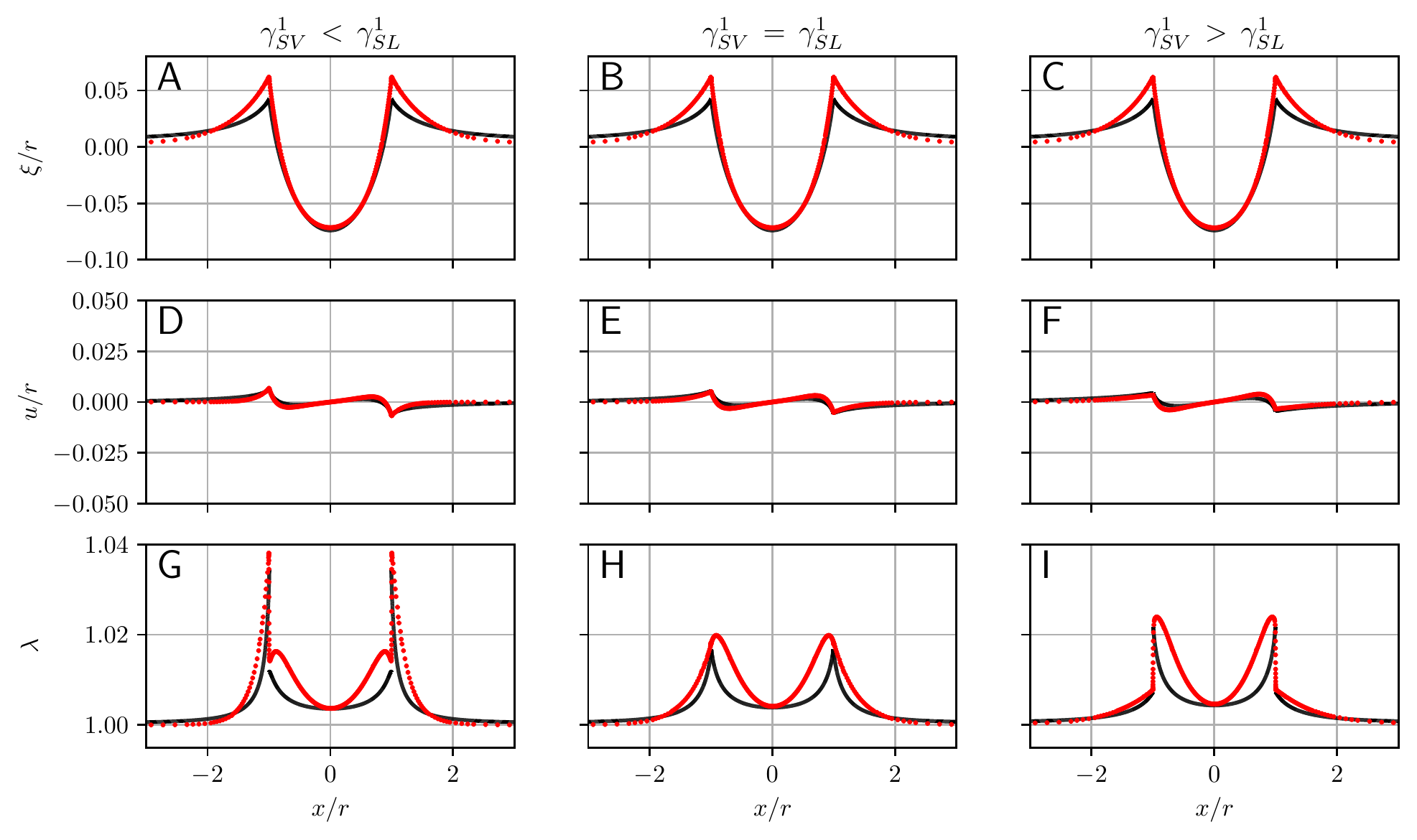}}
\caption{Same as in Fig.~\ref{fig:deformations}, but for $\lambda=1$. In this case, the Shuttleworth effect is small in comparison to that for $\lambda=1.2$, leading to much weaker horizontal displacements $u$ and smaller change in surface stretch $\lambda$ as compared to Fig.~\ref{fig:deformations}.}
\label{fig:deformations_lambda0}
\end{figure}

For completeness, in Fig.~\ref{fig:deformations_lambda0} we also report the data for the case without prestretching of the substrate, i.e., for $\lambda=1$. In comparison to the prestretched case ($\lambda=1.2$) in Fig.~\ref{fig:deformations}, the Shuttleworth effect is much weaker. The reason for this is that our choice for the function $g(\lambda)$ that governs the stretch-dependence of the surface energy exhibits a minimum at $\lambda=1$. Owing to the weak Shuttleworth effect, the horizontal displacements in Fig.~\ref{fig:deformations_lambda0} are much smaller than those in Fig.~\ref{fig:deformations}. Similarly, the surface-stretch $\lambda(x)$ exhibits only small variations along the interface. Note finally that there is no qualitative change anymore between the cases of different asymmetry.

\section{Discussion}\label{sec:discussion}

In summary, we have investigated the static wetting behavior of drops on elastic substrates in the presence of the Shuttleworth effect. We have presented two rather different models: a macroscopic one admitting a detailed description of large-deformation elasticity, and a mesoscopic one offering the possibility of extensions to dynamics and multiple drops. Below we summarise the implications of our work, from the experimental perspective and from the modeling perspective.

A central finding is that the influence of the Shuttleworth effect depends strongly on whether the strain-dependence of the surface energy is symmetric or asymmetric between the ``wet'' and ``dry'' parts of the substrate. The most prominent aspect that is governed by the Shuttleworth effect pertains to the horizontal displacements below the contact line. When the Shuttleworth effect is strongly asymmetric ($\frac{\partial \gamma_{SV}}{\partial \lambda} \neq \frac{\partial \gamma_{SL}}{\partial \lambda}$), 
significant horizontal displacements appear oriented to the side where the Shuttleworth effect is largest. By contrast, for a symmetric Shuttleworth effect ($\frac{\partial \gamma_{SV}}{\partial \lambda} = \frac{\partial \gamma_{SL}}{\partial \lambda}$), the horizontal displacements remain much smaller than the typical vertical displacements. A similar conclusion was already drawn in the limiting case of stiff substrates \cite{WeAS2013sm,AnSn2016el}, for which a tangential force $\frac{\partial \gamma_{SL}}{\partial \lambda} - \frac{\partial \gamma_{SV}}{\partial \lambda}$ was found to be exerted onto the elastic layer. Our results generalise this observation for substrates of arbitrary softness, including the possibility of large elastic deformations. We remark that very large tangential displacements were recently observed for wetting of drops on hydrogels \cite{KimStyPRX2021}. In that case, however, there was also a strong contact angle hysteresis. The pinning of the contact line leads to additional pinning forces that can enhance/reduce the horizontal displacements. Importantly, our findings show that strong horizontal displacements can persist at equilibrium, in the absence of pinning, when the Shuttleworth effect is strongly asymmetric.

Both symmetric and asymmetric Shuttleworth effects have been reported in experiments that explore the dependence of the liquid angle on prestretching of the substrate \cite{XJBS2017nc,STSR2018nc,SnRA2018prl}. According to  Young's law, which involves only surface energy differences, the change in liquid angle  directly reflects the asymmetry in the Shuttleworth effect. While Young's law only holds in the limit of rigid substrates, our results confirm that the magnitude and sign of the change in $\theta_L$ with changing prestretch correlates with the  Shuttleworth-asymmetry up to substrates with $\ell_{\rm ec} \lesssim r$; as is typically the case in experiments. This makes the prestretch-induced variation of the liquid angle a powerful tool to assess the Shuttleworth effect. Both symmetric and asymmetric Shuttleworth effect have been indeed reported in experiments on polymeric substrates. A prestretch-independent $\theta_L$ was observed for various types of elastomers \cite{STSR2018nc}. Also for the case of PDMS a strong Shuttleworth effect was inferred by a number of different techniques \cite{XJBS2017nc,SnRA2018prl,BainPRL2021}. From the perspective of physical chemistry, this suggests that the ``surface-elasticity'' that is responsible for the prestretch-dependence is independent of whether or not the substrate is wetted. The case of an asymmetric Shuttleworth effect was observed for glassy polymers \cite{STSR2018nc}. Indeed, the physico-chemical properties that determine the surface energy are quite different in nature as compared to elastomers \cite{STSR2018nc}.

\begin{figure}[H]
\centerline{\includegraphics[width=\textwidth]{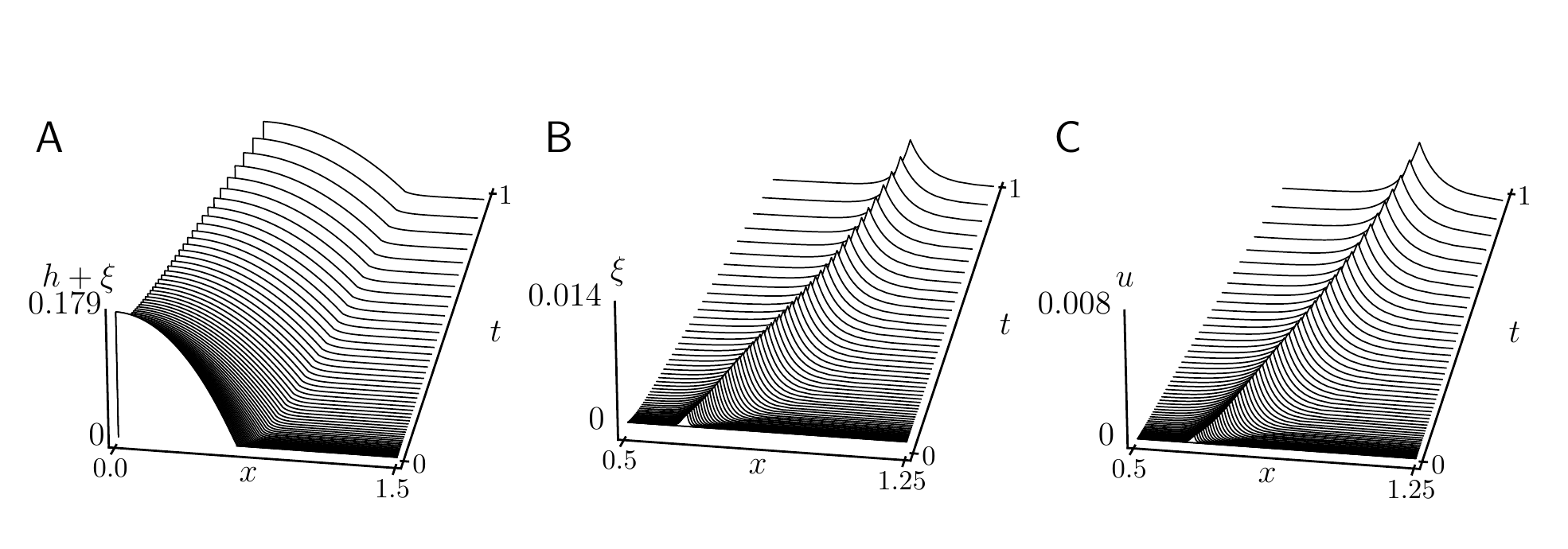}}
\centerline{\includegraphics[width=0.5\textwidth]{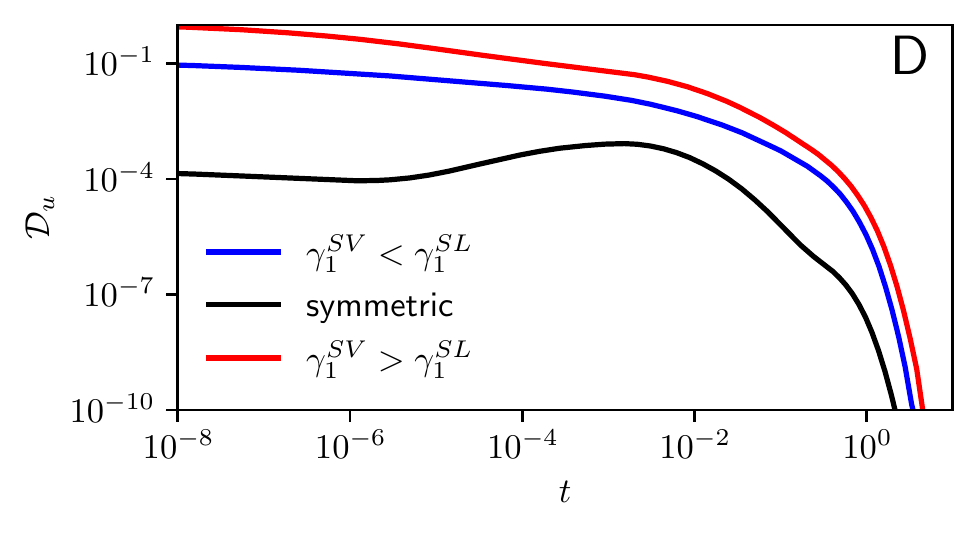}\hspace{0.05\textwidth}}
\caption{The spreading of a droplet as determined by the mesoscopic gradient dynamics model~\eqref{eq:graddyn1}-\eqref{eq:graddyn3}: Panels (a), (b) and (c) present space-time plots for the profiles of film height $h(x,t)$, vertical substrate displacement $\xi(x,t)$, and horizontal substrate displacement $u(x,t)$, respectively in the case of asymmetric Shuttleworth effect with $\gamma_{SV}^1  > \gamma_{SL}^1$. Panel (d) reports the dissipation due to horizontal displacement ${\cal D}_u=\int \frac{1}{\zeta}\left(\frac{\delta\mathcal{F}}{\delta u}\right)^2~\mathrm{d}x$ as a function of time during the spreading. Time and dissipation are given in arbitrary units, while all lengths are given in units of the final drop radius $r$. The elastocapillary length is $\ell_\mathrm{ec}/r\approx5\cdot10^{-2}$ and the Shuttleworth coefficients $\gamma_i^1$ correspond to those of Fig.~\ref{fig:deformations}.}
\label{fig:dynamics}
\end{figure}

From the modeling perspective, we have seen that the gradient dynamics model is able to capture the nontrivial equilibrium features of soft wetting, including the Shuttleworth effect, in spite of its reduced description of elasticity. This validation is very promising as the mesoscopic gradient dynamics model naturally admits dynamical phenomena, such as viscoelastic braking and the Cheerios-effect \cite{HeST2021sm}. To illustrate this perspective, now including the Shuttleworth effect, Fig.~\ref{fig:dynamics} shows some typical dynamical results. They are obtained for a droplet spreading over the substrate towards its equilibrium state. The panels (a-c) show space-time plots of the liquid thickness $h(x,t)$, the vertical displacement $\xi(x,t)$ and the horizontal displacement $u(x,t)$, respectively, for a case with asymmetric Shuttleworth effect. When comparing the dissipation due to horizontal displacements~[Fig.~\ref{fig:dynamics}~(d)], we observe that it is largest for strongly asymmetric Shuttleworth effect; in line with our equilibrium observations. Future investigations using the presented model can demonstrate how the Shuttleworth effect changes dynamical wetting on elastic substrates.

\appendix
\section{Long-wave approximation of mesoscale model}
\label{app-longwave}
The mesoscopic gradient dynamics model obtained in section~\ref{sec:mesomodel-summ} combines an energy functional based on exact metric factors $m(z)=\sqrt{1+z^2}$ and a cubic mobility for the liquid dynamics that can, in analogy to Refs.~\cite{OrDB1997rmp,Thie2007chapter}, be determined via a long-wave approximation of the Navier-Stokes equations. Here, we obtain a long-wave approximation of our dynamical model for the case where all interface slopes are small by expanding the metric factor in the energy functional to $m(z)\approx1 + z^2/2$.
Then, instead of the variations \eqref{eq:varimeso-h}-\eqref{eq:varimeso-u} obtained in the main text, we obtain 
\begin{eqnarray}
\label{eq:vari-lw-1}
\frac{\delta F}{\delta h} &\approx& 
-\gamma_{LV}\left(h''+\xi''\right) +\frac{\partial f}{\partial h}\\
\frac{\delta F}{\delta \xi} &\approx& 
-\gamma_{LV}\left(h''+\xi''\right) 
-\left(\Upsilon'\xi'+\Upsilon\xi''\right)
+\kappa\,\xi \\
\frac{\delta F}{\delta u} &\approx& -\mu' + \kappa\,(u-u'_\infty x)
\end{eqnarray}
where all dashes refer to derivatives w.r.t.\ $x$. 
Further we have \eqref{eq:meso-mu}
\begin{equation}
\label{eq:mu-app}
\mu = \lambda^2 \frac{\partial}{\partial\lambda}\left[\gamma_{SL}(\lambda) + f(h,\lambda)\right].
\end{equation}
with 
\begin{equation}\label{eq:stretch-approx}
\lambda \approx  \frac{1+\frac{1}{2}(\xi')^2}{1-u'}.
\end{equation}
and \eqref{eq:meso-upsilon}
\begin{equation}
\label{eq:ups-app}
\Upsilon =\gamma_{SL}(\lambda) + f(h,\lambda) +  \frac{1}{\lambda}\mu(h,\lambda).
\end{equation}
Introducing \eqref{eq:vari-lw-1}-\eqref{eq:ups-app} into the kinetic equations \eqref{eq:graddyn1}-\eqref{eq:graddyn3} one obtains a consistent mesoscopic gradient dynamics model in long-wave approximation.

Note, however, that the model might be seen as not being asymptotically correct as for small Young angles the interface energy $\gamma_{LV}$ is much larger than the wetting energy $f$ (making the two terms in \eqref{eq:vari-lw-1} the leading balance). Then $\mu$ and $\Upsilon$ each combine terms of different order of magnitude. We argue that nevertheless the much smaller terms in \eqref{eq:mu-app} and \eqref{eq:ups-app} need to be kept as dropping  them would destroy the gradient dynamics structure ensuring thermodynamic consistency. Keeping them also ensures correct long-wave forms of Neumann's law. Also see the related discussion in  \cite{Thie2018csa} and appendix~A of Ref.~\cite{ThAP2016prf}.

\emph{Acknowledgements.}~
We thank Bruno Andreotti, Simon Hartmann and members of SPP~2171 for discussions. We acknowledge financial support from NWO through VICI Grant No. 680-47-632 (to M.H.E.) and an Industrial Partnership Program (a joint research program of Canon Production Printing, Eindhoven University of Technology, University of Twente, and NWO (to E.H.B.). UT and JHS acknowledge support by the Deutsche Forschungsgemeinschaft (DFG) via respective Grants TH781/12 and SN145/1-1 within SPP~2171.

\end{document}